\documentclass[aps,pre,reprint,showpacs,superscriptaddress]{revtex4-1}
\usepackage{graphicx}
\usepackage{dcolumn}
\usepackage{bm}
\usepackage{color}
\usepackage{lipsum}
\usepackage{mathrsfs}
\usepackage{lineno}
\usepackage{siunitx}
\usepackage{dcolumn}
\usepackage{amsmath}
\usepackage{hyperref}
\hypersetup{hypertex=true,colorlinks=true,linkcolor=blue,anchorcolor=blue,citecolor=blue,urlcolor=blue}
\usepackage{enumerate}
\usepackage{multirow}
\usepackage{appendix}
\begin{document}
\title{A high-efficiency proton-boron fusion scheme taking into account the effects of quantum degeneracy}
\author{S. J. Liu}
\affiliation{Institute for Fusion Theory and Simulation, School of Physics, Zhejiang University, Hangzhou, 310058, China}
\author{D. Wu}
\email{dwu.phys@sjtu.edu.cn}
\affiliation{Key Laboratory for Laser Plasmas and School of Physics and Astronomy, and Collaborative Innovation Center of IFSA (CICIFSA), Shanghai Jiao Tong University, Shanghai, 200240, China}
\author{T. X. Hu}
\affiliation{Institute for Fusion Theory and Simulation, School of Physics, Zhejiang University, Hangzhou, 310058, China}
\author{T. Y. Liang}
\affiliation{Institute for Fusion Theory and Simulation, School of Physics, Zhejiang University, Hangzhou, 310058, China}
\author{X. C. Ning}
\affiliation{Institute for Fusion Theory and Simulation, School of Physics, Zhejiang University, Hangzhou, 310058, China}
\author{J. H. Liang}
\affiliation{Key Laboratory for Laser Plasmas and School of Physics and Astronomy, and Collaborative Innovation Center of IFSA (CICIFSA), Shanghai Jiao Tong University, Shanghai, 200240, China}
\author{Y. C. Liu}
\affiliation{Institute for Fusion Theory and Simulation, School of Physics, Zhejiang University, Hangzhou, 310058, China}
\author{P. Liu}
\affiliation{Institute for Fusion Theory and Simulation, School of Physics, Zhejiang University, Hangzhou, 310058, China}
\author{X. Liu}
\affiliation{Key Laboratory for Laser Plasmas and School of Physics and Astronomy, and Collaborative Innovation Center of IFSA (CICIFSA), Shanghai Jiao Tong University, Shanghai, 200240, China}
\author{Z. M. Sheng}
\email{zmsheng@zju.edu.cn}
\affiliation{Institute for Fusion Theory and Simulation, School of Physics, Zhejiang University, Hangzhou, 310058, China}
\author{Y. T. Zhao}
\affiliation{MOE Key Laboratory for Nonequilibrium Synthesis and Modulation of Condensed Matter, School of Physics, Xi’an Jiaotong University, Xi’an 710049, China}
\author{D. H. H. Hoffmann}
\affiliation{MOE Key Laboratory for Nonequilibrium Synthesis and Modulation of Condensed Matter, School of Physics, Xi’an Jiaotong University, Xi’an 710049, China}
\author{X. T. He}
\affiliation{Institute for Fusion Theory and Simulation, School of Physics, Zhejiang University, Hangzhou, 310058, China}
\author{J. Zhang}
\affiliation{Key Laboratory for Laser Plasmas and School of Physics and Astronomy, and Collaborative Innovation Center of IFSA (CICIFSA), Shanghai Jiao Tong University, Shanghai, 200240, China}
\affiliation{Beijing National Laboratory for Condensed Matter Physics, Institute of Physics, Chinese Academy of Sciences, Beijing 100190, China}
\date{\today}

\pacs{}

\begin{abstract}
The proton-boron (p-$^{11}$B) reaction is regarded as the holy grail of advanced fusion fuels, since the primary reaction produces three $\alpha$ particles with few neutrons and induced radio-activities from second order reactions.
Compared to the Deuterium-Tritium reaction a much higher reaction temperature is required. 
Moreover, bremsstrahlung energy losses due to the high nuclear charge of boron deem it seemingly apparent than a fusion reactor based on Deuterium-Tritium plasma in equilibrium is to say the least very difficult.
It is becoming more appealing to collide intense laser beams or accelerated proton beams with a boron target to produce p-$^{11}$B reactions. 
The fusion yield of p-$^{11}$B reactions is closely related to proton beam parameters and boron target conditions such as density, temperature, and ingredients. 
Quantum degeneracy will increase fusion yields by reducing the stopping power of injected protons. 
In this work, we suggest a high-efficiency scheme for beam-target p-$^{11}$B fusions via injecting a MeV proton beam into a highly compressed quantum degenerated boron target.
Such a boron target can be achieved via quasi-isentropic compression of solid boron by using precisely shaped laser pulses. 
Our results indicate that for densities ranging from $10^3$ to $10^4\rho_s$, where $\rho_s$ is the density of solid boron, 
contributions of bound and free electrons to the stopping of protons can be completely disregarded and dramatically reduced respectively.
The result is an increase in fusion yield by orders of magnitude.
Furthermore, in order to achieve multiplication factor $F$ greater than one, with $F$ defined as the ratio of output fusion energy to the energy of injected protons, 
it is found there exits a minimum possible density of boron target, which is $2.15 \times 10^4 \rho_s$ when the kinetic energy of injected protons is $0.8$ MeV.
\end{abstract}

\maketitle

\section{Introduction}
Apart from the advances in fusion research that made the headlines in public non-scientific journals, such as the record in confinement time at high temperature of the Hefei Experimental Advanced Superconducting Tokamak (Xinhua  Updated: 2021-12-31 17:14 May 28, 2021), the 59 Megajoules of fusion power reported from the Joint European Torus in a Deuterium Tritium fusion experiment (JET, 2021 FAZ 2022-02-09), and the burning plasma in an inertial fusion experiment  from the National Ignition Facility of the Lawrence Livermore National Laboratory ( LLNL, 2021; New York Times 2021-08-17), there was a great number of more quiet achievements that demonstrate an accelerated pace towards the final goal of fusion energy \cite{Tikhonchuk2021,Gus2021,Liu2021,Weber2021}.
While the mainstream of research and technological development is directed towards the Deuterium-Tritium reaction for fusion energy, the $^{11}$B(p,$\alpha$)2$\alpha$ process got renewed attention \cite{Hora2021,Ribeyre2022,Margarone2022}, since non-equilibrium conditions, as they are available in laser generated plasma, may turn out to be favorable to enhance the fusion yield \cite{Labaune2013}. The cross-section of the $^{11}$B(p,$\alpha$)2$\alpha$ together with reactions like $^{12}$C(e,e’p)$^{11}$B provides a direct probe for nuclear structure properties of $^{12}$C \cite{Hoffmann1981,Calarco1984,Allas1964,Segel1965}.
Moreover, the abundance of $^{11}$B in the universe is still an unresolved problem.
Therefore, the investigation of these reactions will also contribute to solve the mystery of the low astrophysical abundances of the light elements Li, Be, and B in young main-sequence F and G stars \cite{Lamia2011,Hoffmann1983}.
Besides, the $^{11}$B(p,$\alpha$)2$\alpha$ reaction provides a new method of cancer treatment \cite{Cirrone2018}. 

Despite the recent progress, there is still a long way to go until fusion energy will finally be the solution  the global energy problem.
All routes to fusion energy, as there are magnetic confinement fusion and inertial fusion carry different inherent problems and at the moment it is not clear where the chances of success are highest. Therefore, all possible routes should presently be investigated thoroughly.
However, there are not only technical problems or unsolved physics details, there is also the problem of supply of fusion fuel, especially Tritium in case of the DT-fusion reaction.
The start of the ITER reactor in the 2030s will use the world’s stockpile of tritium and it will take a while until tritium may be supplied  by using fusion neutrons and lithium within the reactor.
Therefore, we deem it important to investigate alternative routes using the $^{11}$B(p,$\alpha$)2$\alpha$ reaction.
This reaction  is a good candidate since the  boron is readily available and neither in the entrance channel nor in the exit channel of the reaction radioactivity is involved \cite{Liu2021}.
There are of course reactions of second order, where neutrons are present and induced radioactivity may occur, but on a low and acceptable level \cite{Weber2021}.
Therefore, at government level in Europe a discussion started to license fusion power plants outside of the restrictions that are valid for fission power plants because it releases few radioactive products (Markus Roth at Fusion Day at GSI Darmstadt Nov. 10$^{\text{th}}$, 2022).
But there is a price to pay. The physical conditions to be met for fusion power based on proton boron fusion are much more demanding than those of DT. 
On the one hand, p11B requires much higher ignition temperature as compared to the D-T reaction, and on the other hand radiation losses due to bremsstrahlung are overwhelmingly high and thus prevent a burning plasma in equilibrium conditions and equimolar fuel constituents, resulting eventually in no or very little net energy output \cite{Moreau1977, Putvinski2019}. However, with advances in high-intensity laser technology, the laser-induced proton-boron beam target nuclear fusion is gradually becoming more attractive.

\begin{figure*}[htbp]
	\includegraphics[width=1.0\textwidth]{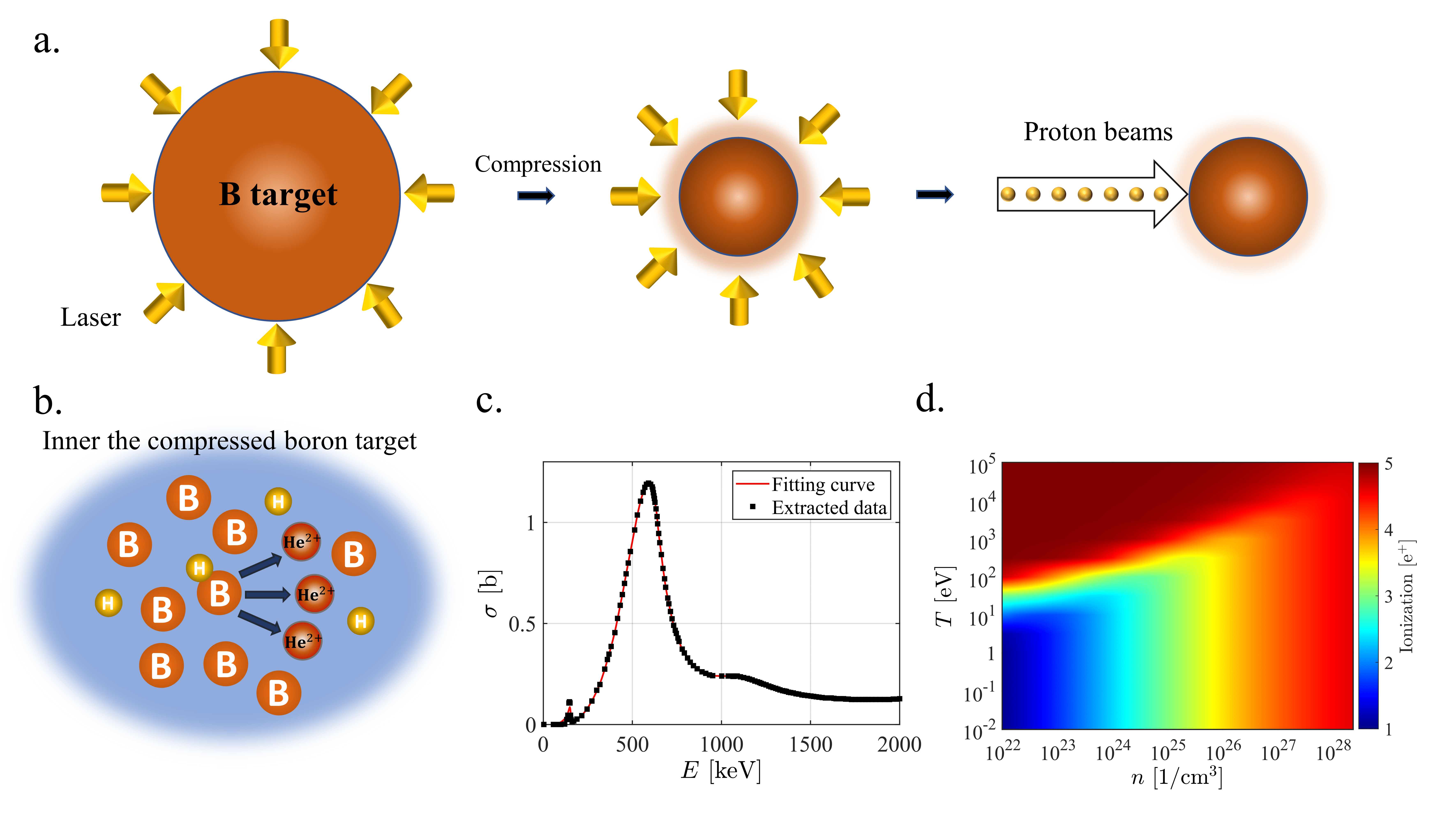}
	\caption{(a)-(b) Diagram of a high efficiency p-$^{11}$B fusion scheme, (c) p-$^{11}$B fusion cross section as a function of center-of-mass energy, where the data of black square is extracted from the work of Nevins and Swain \cite{Nevins2000}, (d) the ionization degree of boron target as a function of temperature and density of boron target.}
	\label{Fig:1}
\end{figure*}

Based on this idea, a number of groups \cite{Belyaev2005, Labaune2016, Kimura2009, Giuffrida2020, Labaune2013} have performed a series of experiments on p-$^{11}$B fusion reactions and measured the yields of $\alpha$ particles. The yields have been increased from about $10^5\ \mathrm{sr}^{-1}$ in 2005 \cite{Belyaev2005, Kimura2009} to about $10^{10}\  \mathrm{sr}^{-1}$ in 2020 \cite{Giuffrida2020}. 
Giuffrida et al. \cite{Giuffrida2020} have investigated the p-$^{11}$B beam-target fusion reactions and calculated the fusion yields. However, the stopping process of protons in detail has not been analyzed and there, thus still leaving open ambiguities involving the interaction between the intense proton beams and the boron target. This interaction depends largely on the intensity of proton beams and the conditions of the boron target such as temperature, density, composition and others.
Beam intensity is influencing the stopping process as has recently been demonstrated \cite{Ren2020}.
The same is obviously true for the target and plasma parameters.
It is important to uncover the relationship between these factors and their respective influence on the reaction probability.
Among these factors, the degree of degeneracy is defined as
$\Theta = {T_F}/{T_e}$, with $T_F$ and $T_e$ representing the Fermi energy and thermal temperature, respectively. 
Electrons have to obey Fermi-Dirac statistics. For $\Theta\ll1$ Boltzmann statistics may be applied \cite{Zwicknagel1999}.

In this paper, the quantitative relationship between the yields of p-$^{11}$B beam-target nuclear reactions and the density of boron targets is derived. 
We find that the effect of quantum degeneracy will increase the fusion yields mediated by the effect of reduced the stopping power of the beam protons \cite{Ningxc2022}.
Based on that, we suggest a high efficiency scheme for beam-target p-$^{11}$B fusions, where a MeV proton beam is injected into a highly compressed quantum degenerated boron target.
An outline of the suggested scheme is displayed in Fig.\ \ref{Fig:1}. 
A highly compressed quantum degenerated boron target can be achieved via quasi-isentropic compression of a solid boron by using precisely shaped laser pulses. 
Our results indicate that for boron target of densities ranging from $10^3$ to $10^4\rho_s$, 
contributions of bound and free electrons to the stopping of protons can be completely disregarded and dramatically reduced respectively, which therefore results in orders of magnitudes increment of fusion yields. 
Furthermore, in order to achieve multiplication factor $F$ greater than one, we find that there exits a minimum required boron target density of $2.15 \times 10^4 \rho_s$ when the kinetic energy of injected protons is $0.8$ MeV.
 
The structure of this paper is organized as follows. In section II, a quantitative relationship between the reaction yields of p-$^{11}$B beam-target nuclear reactions as a function of proton stopping power per unit density is derived.  In section III, contributions from free electrons, bound electrons and ions to the stopping power of protons are analyzed and compared with PIC simulations. 
In section IV, in order to achieve multiplication with $F>1$, the relation between the minimum possible density of compressed boron and the kinetic energy of injected protons is analyzed.
Finally, conclusions and discussions are displayed in section V. 

\section{Beam-target fusion yields}
In general, to calculate the reaction yields of the proton-boron nuclear fusion, we first need to integrate the relative velocity distribution according to the cross section of the proton-boron nuclear fusion under the center of mass system to get the average reaction rate \cite{Atzeni2004}
\begin{equation}
\langle \sigma v \rangle=\int_{0}^{\infty} \sigma(v)vf(v)dv, \label{sigma}
\end{equation}
where $f(v)$ is the distribution function of the relative velocities of protons to boron nuclei, and $\sigma(v)$ is the corresponding cross section with $v$, which is the relative velocity of protons to boron nuclei. 
The number of reactions per unit time per unit volume, namely the volumetric reaction rate, is then calculated as $R=n_p n \langle\sigma {v}\rangle$,
where $n_p$ is the number density of protons, and $n$ is the number density of boron nuclei. 
It shows that the volumetric reaction rate is proportional to the density of protons and boron nuclei \cite{Atzeni2004}. 
Finally, the total reaction number under a certain volume in a certain energy confinement time is obtained by multiplying $R$ with the total volume and confinement time. 

However, this method is strictly applicable only when the proton boron plasma is in thermal equilibrium. For non-equilibrium states, such as the process of projectile-target interaction, the relative velocity distribution changes rapidly. 
Therefore, the average reaction rate also changes with time.
Here we propose a simple model to calculate the nuclear yields of beam-target reactions.
In this model, we just consider the process of projectile-target interaction in p-$^{11}$B beam-target fusion while neglecting thermonuclear contributions.
The previous work of Giuffrida \cite{Giuffrida2020} also proved that thermonuclear contributions to the fusion yields are negligible in the beam-target process.
The density of incident protons is considered to be low, therefore the action of the proton beam on the boron target can be regarded as small perturbation. The temperature of the proton beam and boron target are kept at low levels, meaning that the relative velocities of protons and boron nuclei equal the injected velocities of the proton beam.
Due to the interaction of the proton beam and boron target, when the protons move inside the boron target, their velocities will be decelerated gradually.

We assume a small cloud of protons to be injected into the boron target, and the proton number in this cloud is $N_p$. According to the proton energy, the deceleration process of this cloud inside of the boron target can be divided into many segments with infinitesimal volume, and the energy of protons is considered constant within a respective segment. In the $i$-th segment, the density of protons $n_{pi}$ and the volume $V_i$  will satisfy $N_p=n_{pi} V_i$. We only need to calculate the reaction number of every segment and sum them up. Then we will get the total reaction number during the deceleration process of protons.

During the deceleration process, the energy of protons has slowed down from $E_{pi}$ to $E_{pi}-\delta E_p$ in laboratory coordinates in the $i$-th segment. 
The average reaction rate in the $i$-th segment becomes $\sigma (v_i) v_i$, and then we can express the volumetric reaction rate in the $i$-th segment as $R_i=n_{pi} n \sigma (v_i) v_i$.
The reaction number in the $i$-th segment during the time internal $\delta t_i$ is $R_i \delta t_iV_i=nN_p\sigma ({v_i}) v_i\delta t_i=nN_p\sigma (E_{pi}) \delta z_i$, 
where $\delta t_i$ and $\delta z_i$ are the deceleration time and the deceleration distance of the proton cloud in the $i$-th segment respectively, satisfying $\delta z_i=v_i\delta t_i$.
We have
\begin{equation}
	\delta z_i=\int_{E_{pi}}^{E_{pi}-\delta E_p} \frac{1}{dE/dx}dE, \label{deltaz}
\end{equation}
where $dE/dx=-S$, and $S$ is the stopping power of protons, which is also called the stopping force.

By integrating all segments, we can obtain the total reaction number as
\begin{equation}
R_T = n N_p \int_{0}^{E_p} \frac{\sigma (E)}{S}dE, \label{R_total}
\end{equation}
where $\sigma(E)$ is the p-$^{11}$B fusion cross section as a function of center-of-mass energy, as is shown in Fig.\ \ref{Fig:1}(c). The data of black square is extracted from the work of Nevins and Swain \cite{Nevins2000}. Eq.\ (\ref{R_total}) elucidates the relationship between the number of reactions and the energy loss of protons in a boron target, which can be written into the form,  
\begin{equation}
R_T = N_p P = N_p\int_{0}^{E_{p}} \frac{\sigma (E)}{S/n}dE, \label{RP}
\end{equation}
where $P$ is the rate of a single proton triggering p-$^{11}$B nuclear reactions during the whole deceleration process, and $E_p$ is the initial energy of injected protons.
A similar formula as Eq.\ (\ref{RP}) had also been derived by Giuffrida \cite{Giuffrida2020}.

It is worth noticing that since the cross section of the p-$^{11}$B fusion can not be changed, $P$ only depends on the stopping power per unit density $S_a=S/n$. 
Obviously, $S_a$ is fully determined by the thermodynamic state (density and temperature) of the boron target \cite{Skupsky1977, Fano1963, Clauser2018, Hayes2020}.

\section{The Stopping power of protons}
Contributions to the stopping power of injected protons can be divided into three parts, i.e., stopping by collisions with free electrons, bound electrons and nuclei.
Generally speaking, the contribution from nuclear stopping is small unless at low incident energy. 
However, at highly compressed and quantum degenerate plasmas, the nuclear contribution can not be ignored easily before a detailed analysis of the specific situation. 
In order to comprise all the three contributions, we here take the stopping power of protons as,
\begin{equation}
S = S _f + S_b + S_n, \label{S}
\end{equation}
where $S_f$ is from free electrons, $S_b$ is from bound electrons, and $S_n$ is from nuclei. 
As free and bound electrons contribute separately, to distinguish them, 
the ionization degree of boron target with given density and temperature needs to be determined.
According to the shielded hydrogen model and the single electron counting model \cite{Heltemes2012}, the ionization degree as a function of density and temperature is displayed in Fig.\ \ref{Fig:1} (d).
For highly compressed and quantum degenerated  (low temperature) plasmas, the degree of ionization depends only on the density.
In our following analysis, the initial ionization state of boron target is chosen according to Fig.\ \ref{Fig:1} (d).

\subsection{The free electronic stopping power}

The stopping contribution from free electrons is closely related to quantum degeneracy. 
To indicate the effect of degeneracy, here both theoretical analysis and computer simulations are displayed and compared to each other. 
We divide the part of theoretical analysis into semi-classical and quantum parts respectively. 
Simulations are also divided into two parts: the classical part, and the quantum part considering both Fermi-Dirac statistics and the Pauli exclusion principle.

As for the stopping contribution from free electrons, it was intensively analyzed with both the semi-classical partial wave scattering (SPWS) method \cite{Sigmund1982, Arista2007, Clauser2013, Clauser2018} and the dielectric function method \cite{Jose2014, Lindhard1964, Maynard1982, Maynard1985, Skupsky1977}. Within the dielectric formalism, the free electronic stopping power of a bare ion of mass $m_b\gg m_e$ and charge $Z_be$ ($m_e$ and $e$ are the electron mass and the elementary charge, respectively) moving with velocity $v_p$ is given by \cite{Maynard1985, Jose2014,Deutsch2016},
\begin{equation}
	S_f = \frac{2}{\pi} \left ( \frac{Z_be}{v_p} \right ) ^ 2 \int_{0}^{\infty} \frac{dk}{k} \int_{0}^{kv_p}d \omega \omega \operatorname{Im} \left ( \frac{-1}{\varepsilon (\omega,k)} \right ), \label{Sf}
\end{equation}
where $\varepsilon(k,\omega)$ is the complex dielectric function of the medium which depends on the wave number $k$ and angular frequency $\omega$ of the electromagnetic disturbance caused by the bypassing projectile,
\begin{equation}
	\varepsilon(\omega,\boldsymbol{k})=1+\frac{4\pi}{k^2}\Pi_0(\omega,\boldsymbol{k})
\end{equation}
where $\Pi_0(\omega,\boldsymbol{k})$ is the free-electron density response function. The proportionality of $S_f$ with $Z_b^2$ is a signature of linear-response theory.

\begin{figure*}[htbp]
	\includegraphics[width=1.0\textwidth]{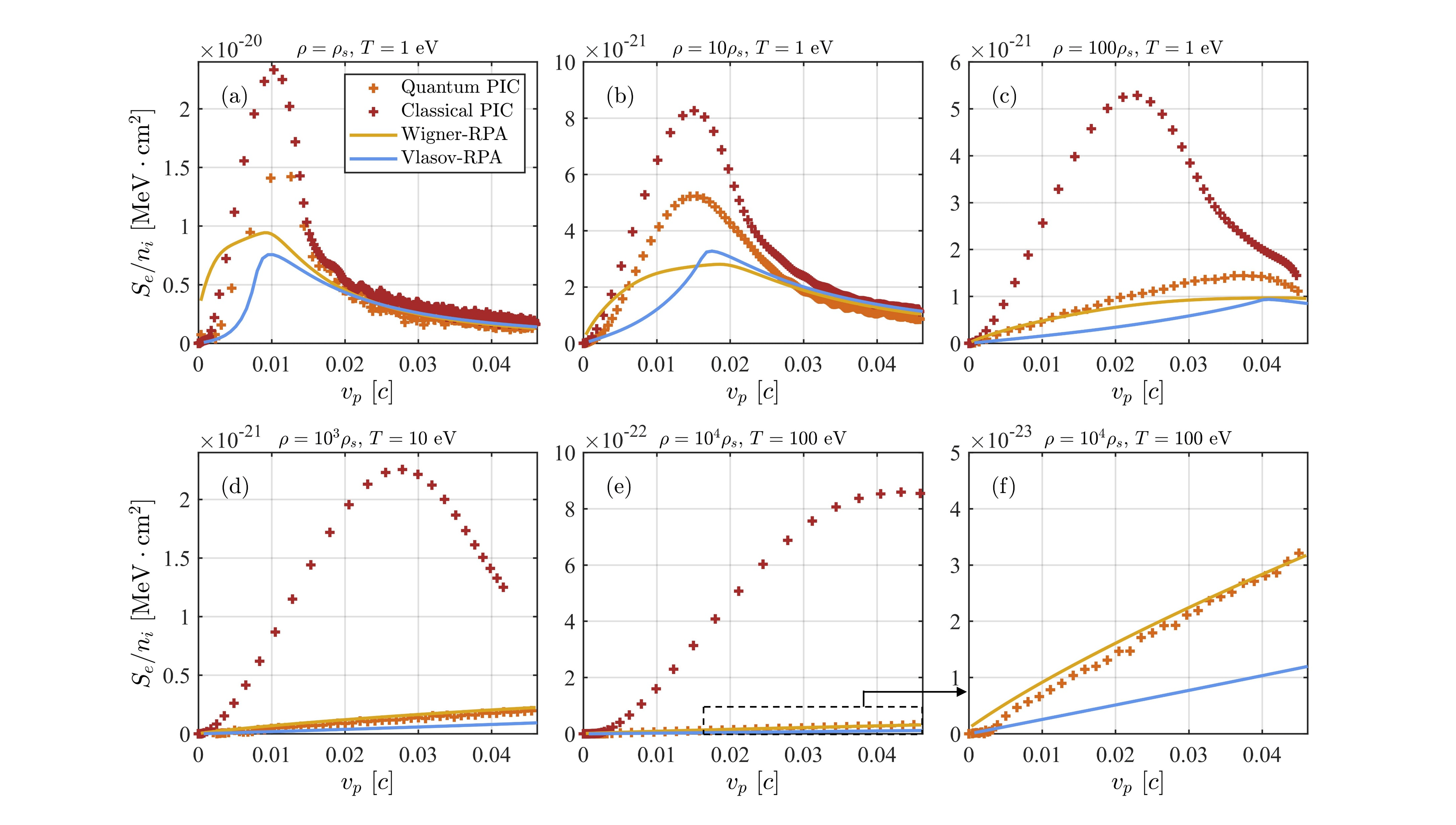}
	\caption{The stopping power per unit density from free electrons as a function of densities and injected velocities of protons, where analytical results and values extracted from simulation by LAPINS code are compared with each other, in which the analytical curves including `Wigner-Poisson' and the semi-classical `Vlasov-Poisson' methods, 
	and the numerical curves including values extracted from `Quantum PIC' and `Classical PIC' simulations.}
	\label{Fig:2}
\end{figure*}
Semi-classically, the stopping power of free electrons that dismisses quantum wave effect can be derived from the Vlasov-Poisson equations,
\begin{equation}
	\label{Vlasov}
		(\partial_t +  \frac{\boldsymbol{p}}{m_e} \cdot \partial_r)f(\boldsymbol{r},\boldsymbol{p},t)+ q\nabla \phi \cdot \partial_{\boldsymbol p}f(\boldsymbol{r},\boldsymbol{p},t) =0,
\end{equation}
and
\begin{equation}
		\nabla ^2 \phi = -4\pi e \left [ Z_b\delta (\boldsymbol{r}-\boldsymbol{v}_pt)- \frac{1}{m_e}\int d\boldsymbol{p}f+ Zn \right ].
\end{equation}
The $\delta$ function stands for the projectile ion moving with velocity $\boldsymbol{v}_p$. The last term in the Poisson equation represents the static plasma ion background.
The $f$ involved in the Vlasov-Poisson equations is Fermi-Dirac distribution. The semi-classical response function is
\begin{equation}
	\Pi_0(\omega,\boldsymbol{k})_{s.c}= -\int d\boldsymbol{p}\frac{\boldsymbol{k}\cdot \partial_{\boldsymbol{p}}f(\boldsymbol{p})}{\omega-\boldsymbol{k}\cdot \boldsymbol{v}}.
\end{equation}
In quantum analysis, the stopping contribution from free electrons can be derived from the quantum-mechanical dielectric function method, which is also called the random-phase-approximation (RPA) method. With quantum wave effects, Wigner-Poisson equations \cite{Hutx2022,Bertsch1988} are usually used. It is convenient to replace the Vlasov equation with the Wigner equation,
\begin{eqnarray}
(\partial_t +  \frac{\boldsymbol{p}}{m} \cdot \partial_r)f(\boldsymbol{r},\boldsymbol{p},t) &=&\frac{e}{i\hbar}\int d\boldsymbol{\xi} \int \frac{d\boldsymbol{p^\prime}}{(2\pi\hbar)^3}e^{i(\boldsymbol{p^\prime}-\boldsymbol{p})\cdot \boldsymbol{\xi}/\hbar} \\
&\times&[\phi(\boldsymbol{r}+\frac{\boldsymbol{\xi}}{2}) - \phi(\boldsymbol{r}-\frac{\boldsymbol{\xi}}{2}) ]f(\boldsymbol{r}, \boldsymbol{p^\prime},t). \nonumber
\end{eqnarray}
The $f$ involved in the Wigner equation is Fermi-Dirac distribution. In this case the quantum response function is 
\begin{equation}
	\Pi_0(\omega,\boldsymbol{k})_q= -\int d\boldsymbol{p} \frac{f(\boldsymbol{p})-f(\boldsymbol{p}+\hbar \boldsymbol{k})}{\hbar \omega-\boldsymbol{k}\cdot \hbar \boldsymbol{v}-\hbar^2k^2/2m}.
\end{equation}
Taking the response function into $\varepsilon(\omega, k)$ and solving Eq. \ref{Sf} numerically, we will obtain the stopping contribution from free electrons. 
Especially, in the limit of low projectile velocities, $v_p \ll v_{\text{ave}}$, with $v_{\text{ave}}$ the average electrons velocity, $v_{\text{ave}}=(v_{th}^2+v_F^2)^{1/2}$, 
where $v_{th}$ is the thermal velocity of electrons, and $v_F$ is the Fermi velocity, the free electronic stopping power is written as
\begin{equation}
	S_f = \frac{4\pi Z_b^2e^4n_f}{m_ev_F^3}v_p\mathrm{ln}(\Lambda_f),
\end{equation} 
where $n_f$ is the number density of free electrons $n_f=Zn$, and $\mathrm{ln}(\Lambda_f)$ is the Coulomb logarithm,  which changes little with the slowing down of protons.
In the limit of high projectile velocities, $v_p \gg v_{\text{ave}}$, the stopping power of free electrons simplifies to \cite{Jose2014}
\begin{equation}
	S_f = \frac{4\pi Z_b^2e^4}{m_ev_p^2}n_f\mathrm{ln}(\Lambda_f),\label{Sf3}
\end{equation} 
which already provides $1\%$ accuracy for $v_p>2v_{\text{ave}}$. See Appendix A for detailed information.

The statistic model used in the LAPINS code \cite{DWu2019, DWu2017, DWu2021, DWu20192} to deal with the case that disregards quantum degeneracy is based on the classical Boltzmann equation, where the average energy of electrons is only determined by the thermal temperature $T_e$. 
The model used in the LAPINS code to deal with quantum degeneracy is based on the first principle Boltzmann-Uhling-Uhlenbeck (BUU) equation \cite{DWu2020}. 
BUU collisions can ensure that the evolution of degenerate particles is enforced by the Pauli exclusion principle. This principle prevents degenerate particles being scattered into an energy state that is already occupied. By using this code, we have simulated the free electronic stopping power in different densities of boron target. See Appendix B for relevant information.

In Fig.\ \ref{Fig:2}, the stopping power per unit density of free electrons $S_e/n$, as a function of proton velocity is shown. 
The range of proton velocity considered is $0.001-0.05c$, with $c$ the speed of light. 
The densities of boron target in Fig.\ \ref{Fig:2} (a-e) are respectively $\rho_s$, $10 \rho_s$, $100 \rho_s$, $10^3 \rho_s$, $10^4 \rho_s$.

When we compare the cases disregarding and considering quantum degeneracy in Fig.\ \ref{Fig:2}, i.e., ``Classical PIC'' and ``Quantum PIC'', 
we can find the effect of degeneracy does decrease the stopping power per unit density of free electrons $S_e/n$. The reason is due to the following two reasons.
Firstly, as the most probable distribution of electrons is the Fermi-Dirac distribution, the average energy of electrons is much higher than the thermal temperature, which is 
\begin{equation}
T_{\mathrm{eff}}=T_e \left [ \frac{2}{\sqrt{\pi}} F_{1/2}(\eta)(1+e^{-\eta}) \right ]^{2/3},
\end{equation}
where $F_{1/2}(\eta)=\int_{0}^{\infty}x^{1/2}(e^{x-\eta}+1)^{-1}dx$, $\eta=\mu/T_e$, $\mu$ is chemical potential. 
Secondly, the Pauli-exclusion principle ensure that only electrons on the boundary of the Fermi surface contribute to the stopping power, as those electrons deep inside the Fermi surface are frozen.
Therefore, when the density of boron plasma target increases, the stopping power per unit density of free electrons is decreased under the influence of quantum degeneracy.

In the limit of low projectile velocities, the average electron stopping power $S_e/n$ predicted by ``Wigner-RPA'' is proportional to the velocity of protons. In the limit of high projectile velocities, it is inversely proportional to the square of the velocity. It is shown that the results of ``Wigner-RPA'' and ``Quantum PIC'' agree with each other quite well. 
Due to the limitations of semi-classic stopping power theory, results of ``Vlasov-RPA'' departs from ``Wigner-RPA'' and ``Quantum PIC'', but still shows a similar trend. 
Especially, the result of ``Vlasov-RPA'' is well consistent with that of ``Wigner-RPA'' when the density of boron target is lower than 100$\rho_s$. 

\subsection{The bound electronic stopping power}

The model used in both analysis and LAPINS code to calculate the bound electron stopping power is based on the work of Fano \cite{Fano1963}, Trujillo \cite{Cabrera-Trujillo1997} and Gil \cite{JM2020}. Generally, we can write the bound electronic stopping power as
\begin{equation}
S_b = \frac{4 \pi Z_{b}^{2}e^4}{m_e v_p^2} (A-Z) n \mathrm{ln}( \Lambda_b ), \label{Sb}
\end{equation}
where 
\begin{equation}
\mathrm{ln}( \Lambda_b ) \equiv \mathrm{ln} \left [ \frac{2 {\gamma}^2 m_e v_p^2}{\bar{I}(Z,A)} \right ] - {\beta}^2 - \frac{C_K}{A} - \frac{\delta}{2}, \label{lambdab}
\end{equation}
in which $A$ is the atomic number of stopping medium, $Z$ is the ionization degree of the background plasma, $\gamma$ is the relativistic factor of the injected ions, and $\bar{I}(Z,A)$ is the average ionization potential considering the degeneracy effect \cite{Rosmej2021,Hu2017}
\begin{equation}
\bar{I}(Z,A) = U - \bigtriangleup U +T_F, \label{Ibar}
\end{equation}
with $U$ the isolated ionization potential \cite{Thomas1981}, and $\bigtriangleup U$ the ionization potential depression (IPD) \cite{Ciricosta2012}. 
As the density of the boron target is increased, the Fermi energy is also increased, and ionizing the bound electrons needs to overcome an extra energy of $T_F$. 
In Eq.\ (\ref{lambdab}), the latter two terms are related to shell corrections and density effect corrections, respectively. These two terms are based on Fano's original work \cite{Fano1963}, to which the definitions of $C_K / A$ and $\delta / 2$ can be referred.

Fig. \ref{Fig:3} (a) shows the stopping power per unit density of bound electrons $S_b/n$ as a function of proton velocity during the deceleration process in different densities of the boron target. 
It is clear that $S_b/n$ is decreased when the density of the boron target is increased. Especially, it can be completely disregarded when the density of the boron target exceeds $100\rho_s$. 
Moreover, as the density of the boron target is increased, the average ionization potential $\bar{I}(Z,A)$ is increased as well, 
which is also reflected by the increasing peak positions of the bound electronic stopping power per unit density when the density of boron target is increased.

\subsection{The nuclear stopping power}
As for the nuclear stopping power, the typical binary collision method \cite{Sigmund2014} is usually used,
\begin{equation}
S_n = \frac{4\pi Z_b^2 Z_t^2 e^4}{m_tv_p^2}n\mathrm{ln}(\Lambda_n), \label{Sn}
\end{equation}
where $Z_t$ is the nuclear charge of the target particle, $m_t$ is the mass of the target particle, and $\mathrm{ln}(\Lambda_n)$ is the Coulomb logarithm \cite{Starrett2018},
\begin{equation}
\mathrm{ln}(\Lambda_n) =\frac{1}{2} \left [ \mathrm{ln}(1+\frac{b_{\text{max}}^2}{b_{\text{min}}^2})-
\frac{b_{\text{max}}^2/b_{\text{min}}^2}{1+b_{\text{max}}^2/b_{\text{min}}^2} \right ], \label{Ln}
\end{equation}
with
\begin{equation}
b_{\text{min}}=\frac{2Z_bZ_te^2}{M_0v_r^2},\label{bmin}
\end{equation}
in which $M_0=m_bm_t/(m_b+m_t)$ is the reduced mass, $b_{\text{min}}$ represents the closest distance that two charged particles of the same sign, $Z_b$ and $Z_t$, with relative velocity $v_r$, can reach, and $b_{\text{max}}$ is an effective maximum impact parameter for nuclear collisions, with $b_{\text{max}}=max(\lambda_\text{D}, r_{\text{WS}})$. Due to the effects of high density and low temperature, the Debye length $\lambda_\text{D}$ is usually smaller than the Wigner-Seitz radius $r_{\text{WS}} = ({4\pi}/{3n})^{-{1}/{3}}$.

\begin{figure*}[htbp]
	\includegraphics[width=0.9\textwidth]{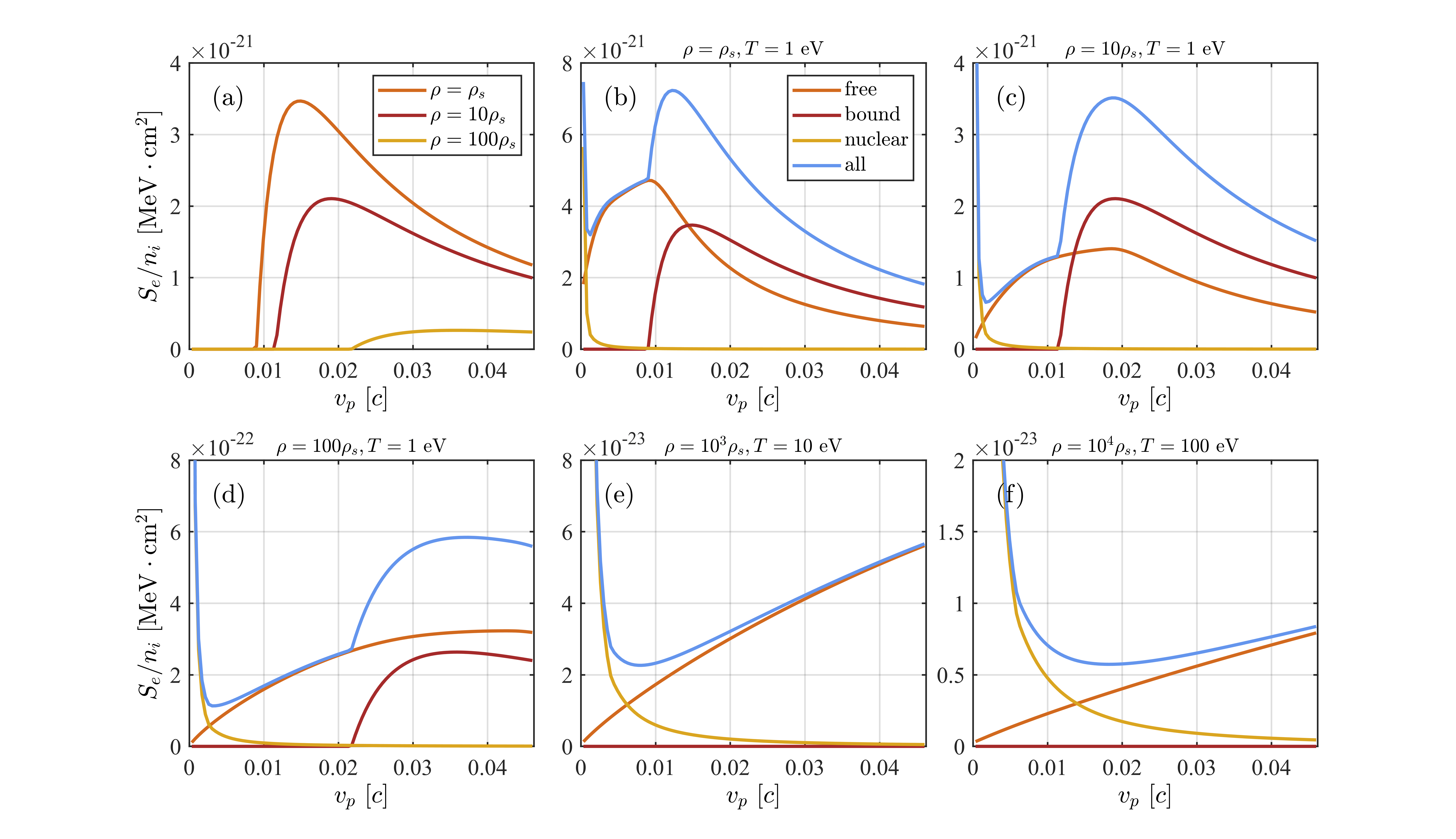}
	\caption{(a) The bound electronic stopping power per unit density as a function of densities and velocities of injected protons. (b)-(e) The stopping power per unit density as a function of densities and velocities of injected protons, with contributions from the total, free electrons, bound electrons, and nuclei plotted, respectively. The label is shown in Fig. (b).}
	\label{Fig:3}
\end{figure*}

Fig.\ \ref{Fig:3} (b-e) show the contributions of all the three components (free electrons, bound electrons, nuclei) to the stopping power per unit density of injected protons in the boron target. As can be seen in Fig.\ \ref{Fig:3}, at the low density region, $S_n/n$ is far smaller than the electronic stopping power per unit density and can be ignored. With the increasing of boron target density, the contribution of nuclei becomes important. Meanwhile, the peak of $S_b/n$ moves to the right and the contribution of bound electrons dwindles. When the target density is $10^3$-$10^4\rho_s$, the contributions of free electrons and nuclei determine the stopping power per unit density, especially $S_e/n$ dominates at high proton velocity while $S_n/n$ dominates at low proton velocity.

\section{Multiplication factor of beam-fusion reactions}

With above analysis, we can easily obtain the total reaction number of  beam-target nuclear fusion
\begin{equation}
R_T = N_pP=N_p \int_{0}^{E_p} \frac{\sigma (E)}{(S_f + S_b + S_n)/n} dE. \label{Rtotal1}
\end{equation}
For the p-$^{11}$B beam-target reaction, the total number of reactions is closely related to the stopping power per unit density of protons in the target. 
As can be seen in Fig.\ \ref{Fig:3}, the total stopping power per unit density $S/n$ decreases when the density of boron target is increased. 
Especially when the density of boron target reaches $10^3\rho_s$, $S_b/n$ almost disappears, the total stopping power per unit density is now determined by
\begin{equation}
	\frac{S}{n} = \frac{4\pi Z_b^2 Z_t^2 e^4}{m_tv_p^2}\mathrm{ln}(\Lambda_n) + \frac{4\pi Z_b^2e^4Z}{m_ev_F^3}v_p\mathrm{ln}(\Lambda_f). \label{cross section}
\end{equation}
The reason for the decrease is due to the following two facts. Firstly, as $b_{\text{max}}$ is inversely proportional to $n^{1/3}$, 
$\mathrm{ln}(\Lambda_n)$ decreases when the density of boron target is increased.
Secondly, as $v_F^3$ is proportional to $n_f=Zn$, the second term of Eq.\ (\ref{cross section}) $\propto \mathrm{ln}(\Lambda_f)/n$ also decreases when the density of boron target is increased. 

\begin{figure*}[htbp]
	\centering
	\includegraphics[width=1.0\textwidth]{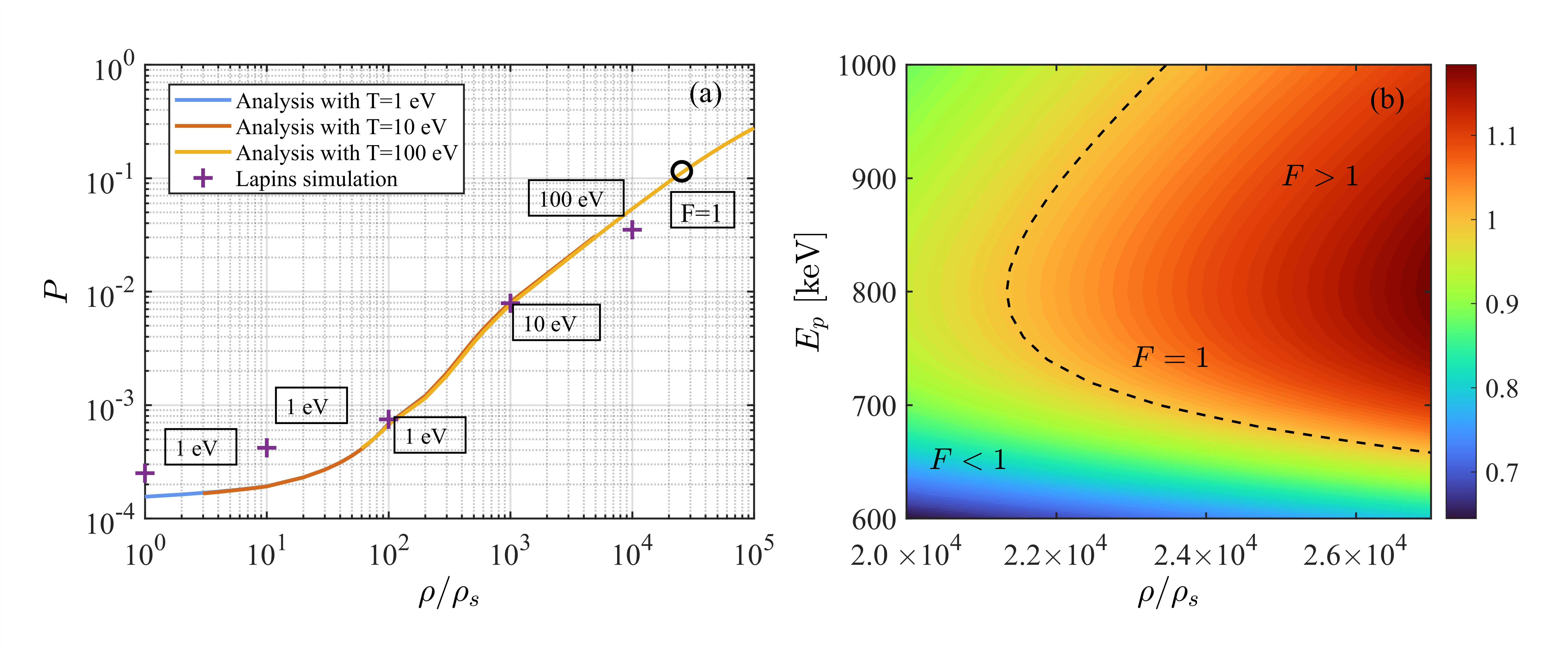}
	\caption{(a) The reaction probability of the p-$^{11}$B nuclear fusion as a function of boron target density. The initial energy of proton beam is 1 MeV. 
		(b) The multiplication factor $F$ of p-$^{11}$B beam-target fusion as a function of boron target density and initial proton kinetic energy. 
		$F$ is defined as the ratio of the fusion energy produced during the deceleration of proton beam to the overall energy of injected protons.}
	\label{Fig:4}
\end{figure*}

Numerically, we have run simulations with different initial densities by using the LAPINS code, with density of $\rho_s$, $10\rho_s$, $100\rho_s$, $10^3\rho_s$, and $10^4\rho_s$, respectively. 
The initial kinetic energy of the proton beam is fixed at $1$ MeV. 
Fig.\ \ref{Fig:4} (a) shows the probability of the p-$^{11}$B beam-target fusion as a function of the boron target density.
The comparisons between theoretical analysis and numerical simulations coincide with each other quite well.
Due to high degree of quantum degeneracy, the effect of temperature is quite small on the reaction probability. 

Quantitatively, it is critical to consider the energy multiplication factor $F$ in order to achieve the net energy gain. Here $F$ is a fundamental quantity in beam-target fusion,
\begin{equation}
F=\frac{P(E_p)Q}{E_p},\label{Ffactor}
\end{equation}
where $E_p$ is the initial proton energy, $Q$ is the fusion $Q$-value. For the p-$^{11}$B fusion reaction, $Q$ equals 8.7 MeV. Theoretical analysis leads to the conclusion that for injected protons of energy of 1 MeV, the threshold density of boron target is $2.35\times 10^4 \rho_s$, beyond which the energy multiplication factor $F$ would be greater than one.

Moreover, the cross section of nuclear fusion depends on the center of mass energy, which is a function of the injected proton energy. Therefore, different injected energy of proton beams may lead to different reaction probabilities. We have calculated the fusion probability with varying injected proton kinetic energies. 
In order to make the $F$ factor more than one, we find when the energy of injected protons is around 800 keV, 
there exists a minimum possible compressed density, which is $2.15\times10^4 \rho_s$, as shown in Fig.\ \ref{Fig:4} (b).

\section{Conclusion}

In conclusion, we suggest a high-efficiency scheme for beam-target p-$^{11}$B fusions via injecting a MeV proton beam into a highly compressed quantum degenerated boron target. The degeneracy effect is found to have an effect on the number of fusion reactions by decreasing the stopping power per unit density of protons in the boron target. At low boron target densities, free electrons and bound electrons dominate the stopping power per unit density. Especially when $E_p=1$ MeV and $\rho=10^3-10^4 \rho_s$, $S_b/n$ and $S_e/n$ can be completely disregarded and dramatically reduced, respectively, which therefore results in orders of magnitude increments in fusion yields. When the injected proton beam has an energy of around 800 keV, there exists a minimum possible compressed density, which is $2.15\times 10^4 \rho_s$ to make the $F$ factor greater than one.

\section{Acknowledgements}
This work was supported by the Strategic Priority Research Program of Chinese Academy of Sciences (Grant No. XDA250050500), the National Natural Science Foundation of China (Grant No. 12075204, No. 11875235, and No. 61627901), and Shanghai Municipal Science and Technology Key Project (No. 22JC1401500). Dong Wu thanks the sponsorship from Yangyang Development Fund.

\begin{appendices}
\renewcommand{\thetable}{\thesection-\arabic{table}}
\renewcommand{\thefigure}{\thesection-\arabic{figure}} 
\renewcommand{\theequation}{\thesection-\arabic{equation}} 
\section{The calculation of the free electron stopping power from Wigner-RPA equations}	
\setcounter{table}{0} 
\setcounter{equation}{0} 
\setcounter{figure}{0}
Consider a homogeneous free electrons gas (FEG) with density $n_f$ and Fermi wave number $k_F = (3\pi^2 n_f)^{1/3}$, the FEG can be characterized by the Lindhard parameter $\chi^2$; these dimensionless quantities are linked to $k_F$ through \cite{Maynard1985, Jose2014}
\begin{equation}
	\chi^2 = (\pi k_F a_0)^{-1}, \label{chi}
\end{equation}
where $a_0 = \hbar ^2 / m_e e^2$ is the Bohr radius ($\hbar$ is the reduced Plank constant) and $\alpha = (4/9\pi)^{1/3}$. It is convenient to replace $k$ and $\omega$ by the dimensionless variables
\begin{equation}
	z=\frac{k}{k_F} \ \text{and} \ u=\frac{\omega}{kv_F}, \label{zu}
\end{equation}
where $v_F=\hbar k_F/m_e$ stands for the Fermi velocity. In this case the dielectric function can be written as \cite{Maynard1985}
\begin{equation}
	\varepsilon(z,u) = 1+\frac{\chi^2}{z^2}[f_1(z,u)+if_2(z,u)], \label{varepsilon}
\end{equation}
with
\begin{equation}
	f_1(u,z) = -\frac{\pi}{8z \Theta}[F(u+z)-F(u-z)]. \label{f1}
\end{equation}
When $\Theta\gg1$, 
\begin{equation}
	F(p)=2p\left [\frac{1}{2}+\frac{1-p^2}{4p}\log\frac{p+1}{p-1} \right ], \label{F(p)}
\end{equation}
and
\begin{equation}
	f_2(u,z) = - \frac{\pi}{8 \Theta z}\log\frac{1+\exp \left ( {\mu}/{T}-\Theta (u+z)^2 \right )}{1+\exp \left ( {\mu}/{T}-\Theta (u-z)^2 \right )}. \label{f2}
\end{equation}
The free electron stopping power can be written in the form
\begin{equation}
	S_f = \frac{4\pi Z_b^2 e^4}{ m_e v_p^2}n_f L_e, \label{Sf1}
\end{equation}
with
\begin{equation}
\begin{aligned}
L_e = \frac{6}{\pi \chi ^ 2} \int_{0}^{v_p / v_F} u du \int_{0}^{\infty} z dz \operatorname{Im} \frac{1}{\varepsilon (z,u)},\label{le}
\end{aligned}
\end{equation}
which depends on the number density and temperature of target nuclei as well as the velocity of protons, satisfying
\begin{equation}
	L_e = \left ( \frac{v_p}{v_r} \right )^3 \mathrm{ln}(\Lambda_f). \label{Le}
\end{equation}
Here $v_r$ is the relative velocity between protons and free electrons, and $v_r=(v_{\text{ave}}^2+v_p^2)^{1/2}$. In the limit of low projectile velocity, $v_p\ll v_{\text{ave}}$,  the Eq. (\ref{Le}) becomes
\begin{equation}
	L_e = \left ( \frac{v_p}{v_r} \right )^3 \mathrm{ln}(\Lambda_f)\simeq \left ( \frac{v_p}{v_F} \right )^3 \mathrm{ln}(\Lambda_f),\label{le1}
\end{equation}
and the stopping power of free electrons simplifies to \cite{Jose2014}
\begin{equation}
	S_f = \frac{4\pi Z_b^2e^4n_f}{m_ev_F^3}v_p\mathrm{ln}(\Lambda_f).\label{Sf2}
\end{equation} 
In the limit of high projectile velocity, $v_p\gg v_{\text{ave}}$, $L_e$ becomes
\begin{equation}
\begin{aligned}
L_e  = \left ( \frac{v_p}{v_r} \right )^3 \mathrm{ln}(\Lambda_f)\simeq \mathrm{ln}(\Lambda_f), \label{Lehigh}
\end{aligned}
\end{equation}
and the stopping power of free electrons simplifies to \cite{Jose2014}
\begin{equation}
	S_f = \frac{4\pi Z_b^2e^4}{m_ev_p^2}n_f\mathrm{ln}(\Lambda_f),\label{Sf2}
\end{equation} 
Note, Eq. (A-12) already gives a one percent accuracy for $v_p>2v_{\text{ave}}$.

\section{Simulation of free electron stopping power with the LAPINS code}
\setcounter{table}{0} 
\setcounter{equation}{0} 
\setcounter{figure}{0}
The statistic model used in the LAPINS code \cite{DWu2019, DWu2017, DWu2021, DWu20192} to deal with the case that disregards quantum degeneracy is based on the classical Boltzmann equation,
\begin{equation}
	\frac{\partial f}{\partial t} + \boldsymbol{v}_k \cdot{\frac{\partial f}{\partial \boldsymbol{r}}} + q_k ( \boldsymbol{E} + \boldsymbol{v}_k \times \boldsymbol{B}) \cdot{\frac{\partial f}{\partial \boldsymbol{p}_t}} = \frac{\partial f}{\partial t} |_{\text{coll}},\label{Boltzmann}
\end{equation}
where the subscript $k$ indicates the species of particles, $f = f(\boldsymbol{r}, \boldsymbol{p}, t)$ is the distribution function, 
\begin{figure}[htbp]
	\centering
	\includegraphics[width=0.4\textwidth]{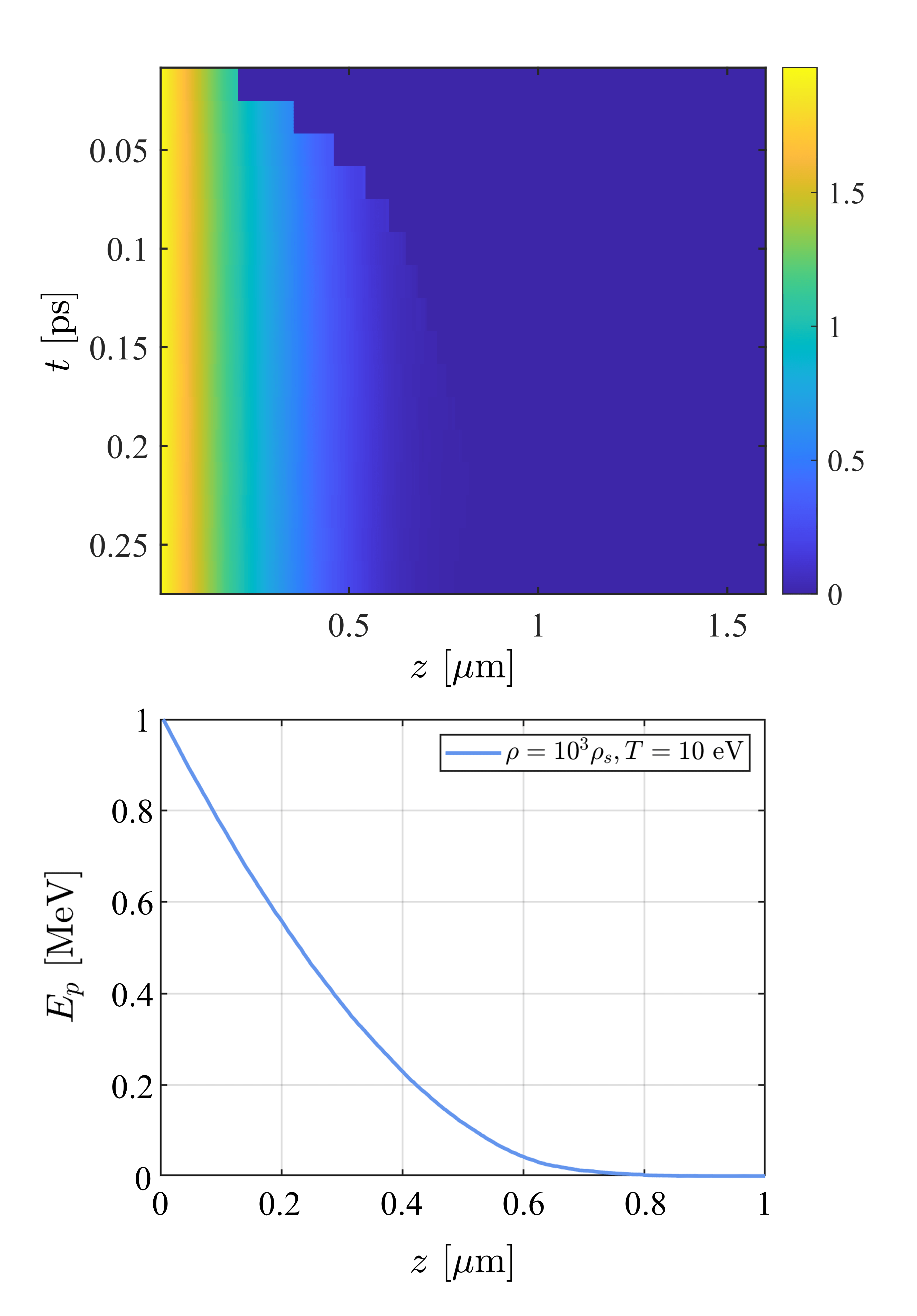}
	\caption{(a) The proton energy as a function of time and incident distance, simulated by the LAPINS code considering Pauli exclusion principle, with boron density of $10^3\rho_s$ and temperature of $10$ eV. 
		The duration of the beam is set long enough to ensure that the beam is still injected when the protons in the front are stopped. We therefore can get stable data describing the proton energy as a function of the incident distance during the deceleration process. (b) The proton energy as a function of the incident distance during the deceleration process.}
	\label{Fig:B1}
\end{figure}
$\boldsymbol{r}$ is the position, $\boldsymbol{p}$ is the momentum, $t$ is the time, $\boldsymbol{v}$ is the velocity, $\boldsymbol{E}$ is the electric field, $\boldsymbol{B}$ is the magnetic field and the collision term ${\partial f}/{\partial t} |_{\text{coll}}$ is Boltzmann collision integral. On this occasion, the average energy of electrons is only determined by thermal temperature $T_e$, and the quantum effect of electrons is completely not considered.

The model used in the LAPINS code to deal with quantum degeneracy is based on the first principle Boltzmann-Uhling-Uhlenbeck (BUU) equation \cite{DWu2020, Domps1997},
\begin{equation}
	\frac{\partial f}{\partial t} + \boldsymbol{v}_k \cdot{\frac{\partial f}{\partial \boldsymbol{r}}} + q_k ( \boldsymbol{E} + \boldsymbol{v}_k \times \boldsymbol{B}) \cdot{\frac{\partial f}{\partial \boldsymbol{p}_t}} = \frac{\partial f}{\partial t} |_{\text{coll}}^{\text{BUU}},\label{BUU}
\end{equation}
where the BUU collisions can ensure that evolution of degenerate particles is enforced by the Pauli exclusion principle. This principle prevents degenerate particles being scattered into an energy state that is already occupied.

In order to give the visual stopping power per unit density of free electrons, we do a series of simulations by the LAPINS code. The initial energy of proton beams is set to 1 MeV. The densities of boron target are respectively $\rho_s$, $10 \rho_s$, $100 \rho_s$, $10^3 \rho_s$, $10^4 \rho_s$, where $\rho_s$ is the density of solid boron target equal to $2.34$ $\mathrm{g/cm ^3}$. The temperatures of the former three cases are set to be 1 eV, while the temperatures of the later two are set to be 10 eV and 100 eV, respectively. The reason why a higher temperature is set for the last two cases is to avoid numerical errors in the integral due to large degeneracy parameter $\Theta$. In fact, when the density of boron target is over 100 $\rho_s$, such changes of temperature will not influence the result a lot. 

For each simulation, the density of proton beam is set low enough to make sure the influence can be regarded as perturbations. Duration of the beam is set long enough to ensure that the beam is still being injected when the velocity of the protons in the front plane is slowed down to zero. 
As shown in Fig.\ \ref{Fig:B1} (a), we can get the proton energy at any position in the target at any time. On this occasion, as is shown in Fig.\ \ref{Fig:B1} (b), we can obtain the data of proton average energy $E_{pi}$ as a function of distance in the deceleration process. We will then get the free electronic stopping power of to protons by $\delta E_{pi}/\delta z_i$. Dividing it by $n$, we can obtain the free electronic stopping power per unit density $S_e/n$.
\end{appendices}

\bibliography{reference}
\end{document}